\documentclass[twocolumn,showpacs,preprintnumbers,amsmath,amssymb,floatfix]{revtex4}

\usepackage{graphicx}
\usepackage{dcolumn}
\usepackage{bm}
\usepackage{color}

\newcommand{\beq}{\begin{equation}}
\newcommand{\eeq}{\end{equation}}
\newcommand{\bey}{\begin{eqnarray}}
\newcommand{\eey}{\end{eqnarray}}

\begin{document}


\title{Stable Gravastars: Guilfoyle's electrically charged solutions}

\author{Ayan Banerjee}
 \email{ayan_7575@yahoo.co.in}
\affiliation {Department of Mathematics, Jadavpur University,
 Kolkata-700032, India.}

\author{J. R. Villanueva}
 \email{jose.villanueva@uv.cl}
\affiliation {Instituto de F\'{\i}sica y Astronom\'{\i}a, Facultad de Ciencias, Universidad de
 	Valpara\'{\i}so, Gran Breta\~na  1111,  Valpara\'{\i}so, Chile.}

\author{ Phongpichit Channuie}
 \email{channuie@gmail.com}
\affiliation {School of Science, Walailak University, Nakhon Si Thammarat, 80160 Thailand.}

\author{Kimet Jusufi}
 \email{kimet.jusufi@unite.edu.mk}
\affiliation {Physics Department, State University of Tetovo, Ilinden Street nn, 1200, Macedonia.}

\date{\today}

\begin{abstract}
Compelling alternatives to black holes, namely, gravitational vacuum star ({\it gravastar}) models, the multilayered structure compact objects, have been proposed to avoid a number of theoretical problems associated with event horizons and singularities. In this work, we construct a spherically symmetric thin-shell charged gravastar model where the vacuum phase transition between the de Sitter interior and the external Reissner--Nordstr$\ddot{\text{o}}$m spacetime (RN) are matched at a junction surface, by using the cut-and-paste procedure. Gravastar solutions are found among the	Guilfoyle exact solutions where the gravitational potential $W^2$ and the electric potential field $\phi$ obey a particularly relation in a simple form $ a\left(b-\epsilon \phi \right)^2 +b_1$, where $a$, $b$ and $b_1$ being arbitrary constants. The simplest ansatz of Guilfoyle's solution is implemented by the following assumption: that the total energy density $8\pi \rho_m+\frac{Q^2}{ r^4}$ = constant, where $Q(r)$ is the electric charge up to a certain radius $r$.
We show that, for certain ranges of the parameters, we can avoid the horizon formation, which allows us to study the linearized spherically symmetric radial 	perturbations around static equilibrium solutions. To lend our solution theoretical support, we also analyze the physical and geometrical properties of gravastar configurations.

\end{abstract}

\keywords{Einstein's field equations; Stellar equilibrium.}

\maketitle

\section{Introduction}
\label{intro}
The final state of the gravitational collapse could lead to the formation of different objects such as neutron stars, white dwarfs, black holes and naked singularities \cite{chandra31,BZ34,landau32,oppen,penrose}. This end-state of collapse is a widely accepted field of research in the scientific community from many perspectives, either theoretical and observational.
However, classical general relativity suffers from some severe theoretical problems. One of them is precisely related to the black hole paradoxical features and naked singularities. In order to resolve these issues, the idea of the existence of compact objects without event horizons has recently been proposed, the so-called {\it gravastar}: an alternative to black holes \cite{Mazur,Mazur(2001),Mazur(2004)}. 
Moreover, some interesting articles have been published. Within these models, a massive star in its final stages could end its life as a gravastar that compresses matter within the gravitational radius $r_s=2GM/c^2$, i.e., very close to the Schwarzschild radius but with no singularity or event horizon.
In this sense, the presence of the quantum vacuum fluctuations are expected to play a non--trivial role at or near the event horizon. Based on the gravastar framework, the solutions of the Mazur and Mottola’s gravastar scenario describe five layers of the gravastar, including two thin--shells. The de Sitter geometry in the interior with an equation of state (EOS) p = -$\rho$,
matches to an exterior Schwarzschild vacuum geometry. Between the interior and exterior geometry, there is a finite--thickness shell comprised of stiff fluid matter with EOS p = +$\rho$. Due to their extreme compactness, however, it would be very difficult to distinguish gravastars from black holes.
As an extension of the Mazur--Mottola model and for physical reasons, Visser and Wiltshire  \cite{Visser} reduced a number of thin shells from 5 layers to 3 layers with a continuous layer of finite thickness, where the phase transition layer was replaced by a single spherical $\delta$-shell. The model, by the definition of the gravastar, a de-Sitter spacetime was matched to a Schwarzschild exterior solution at a junction surface with surface stresses $\sigma$ and $\mathcal{P}$. Moreover, authors provided the full dynamic stability against spherically symmetric perturbations by using the Israel thin shell formalism \cite{Israel} in terms of an effective energy equation. This simplified form of gravastar structure also motivated a number of physicist to
consider different types vacuum geometries in the interior and the exterior. Among these models, Bili\'c and his collaborators have shown how a gravastar structure can  from a Born--Infeld scalar field \cite{Bilic} and the non-linear electrodynamic gravastar \cite{Lobo:2007}. In \cite{DeBenedictis}, gravastar solutions have been studied by replacing the $\delta$--shell with a continuous stress-energy tensor from the asymptotically de Sitter interior to the exterior Schwarzschild solution. Moreover, an electrically charged gravastar has been studied by solving the Einstein--Maxwell field equations in the asymptotically de Sitter interior \cite{Horvat}, whereas a charged gravastar admitting conformality was constructed in \cite{Usmani}. In the same vein, Chan \textit{et al}. \cite{Chan} have studied radiating gravastars by considering Vaidya exterior spacetime. However, it was shown that the interior de Sitter spacetime may also replace by
considering a solution governed by the dark energy equation of state, $\omega=p/\rho$ where $\omega< -1/3$ \cite{Lobo:2005uf}.
Furthermore, much effort has been made to investigate the properties of gravastars in the context of alternative scenarios. These were considered in Refs. \cite{Das,Banerjee,Banerjee:EPJC,Lobo(2013),Ali}.
	
As is well known, not only can an interior regular charged perfect fluid solution not only can contribute to a better understanding of the structure of spacetime, but also physically it also offers many new and interesting solutions. A static solution around a spherically non-rotating charged body was stimulated by the work of Reissner and Nordstr\"om \cite{Reissner}.  Almost at the same time, Hermann Weyl \cite{Weyl,Weyl19} studied vacuum general relativity and electromagnetism, which together established a relation between the metric component $g_{tt}$ and the electric potential $\phi$. A further discussion about this relation appeared in 1947 when Majumdar \cite{Majumdar} generalized this result to the systems without spatial symmetry. Regarding this point of view, further development on Weyl's work was examined much later by Guilfoyle \cite{Guilfoyle}. He considered charged fluid distributions in which the interior of these solutions is characterized by the relation between the gravitational and the electric potential, which are functionally related to each other via $W = W(\phi$). Here $W$ is parametrized by $W^2 = a \left(b-\epsilon \phi \right)^2 +b_1$, where $a$, $b$ and $b_1$ are arbitrary constants, and $\epsilon = \pm$. Specifically,  this relation generalizes the common Weyl relation, when $a = 1$, and the extension of this concept was discussed by Lemos and Zanchin \cite{Lemos(2009)}. Here they obtained a relationship between various fields and matter quantities. The authors also studied quasiblack holes such as frozen stars \cite{Lemos}, and found regular black holes for the Guilfoyle exact solutions \cite{Lemos(2016)} with the discussion of their physical relevances. In addition, Lemos and Zanchin \cite{Lemos:2015wfa} applied the method to study relativistic charged
spheres, and shown that when the central pressure goes to infinity Guilfoyle's
stars also obey the Buchdahl-Andr$\acute{e}$asson bound. Moreover, regarding the Weyl and Weyl--Guilfoyle relations, the general relativistic charged fluids with non--zero pressure turned out to be an important cornerstone for addressing and discovering new solutions. Thus, it is interesting to embark on a study of the Guilfoyle model with the presence of the electrically charged matter. We expect that the gravastar solutions are found within a certain range of the parameters of the model.  
	
The outline of the paper is as follows. In Sec. \ref{struct}, we present Guilfoyle's exact solutions with the ansatz $W^2 = a \left( b-\epsilon \phi \right)^2 +b_1$. In Sec. \ref{spherical}, the basic equations for spherically symmetric spacetime with electrically charged perfect fluid matter distribution are written, satisfying a particular Weyl--Guilfoyle relation, and we discuss the interior and exterior solutions with appropriate boundary conditions in Sec. \ref{gsir}.  In Sec. \ref{rgm}, we briefly review some gravastar models.  In Sec. \ref{cgm}, we present the basic setup for matching the two distinct spacetimes and to obtain models of the thin shell gravastar. We also investigate the constraints on  parameters at the junction interference with a time-like thin shell in Sec. \ref{jc}. Then in Sec. \ref{sg}, we obtain the static gravastar solution and briefly outline the linearized stability analysis with the determined stability regions of the transition layer in Sec. \ref{lsog}. Next, we investigate the stability of gravastar by using the surface mass of the thin shell, and we discuss  in detail some interesting observations. Our conclusions are given in Sec. \ref{sts}. Throughout this work, if not explicitly stated otherwise, we use the units of $G=c=1$.
	
\section{ Structural Equations of Weyl--Guilfoyle charged fluid}
\label{struct}
This section is devoted to presenting the basis equations governing the dynamical behavior of a cold charged gravitating distribution of a relativistic fluid. So, the starting point is to revise the Einstein--Maxwell field equations in a four-dimensional spacetime given by
	
\begin{equation}\label{efe}
G_{\mu\nu}=R_{\mu\nu}-\frac{1}{2}g_{\mu\nu} R = 8\pi\left(T_{\mu\nu} +E_{\mu\nu}\right),
\end{equation}and,
	\begin{eqnarray}\label{fjrelation}
	&\nabla_{\nu}\,F^{\mu\nu} = 4\pi J^{\mu},\\\label{flme}
	&\nabla_{[\alpha}\, F_{\mu \nu]} = 0
	\end{eqnarray}
	where $R_{\mu\nu}$ is the Ricci tensor, $R$ is the Ricci scalar, $T_{\mu\nu}$ is the energy--momentum tensor, for which we assume it is a perfect fluid:
	\begin{equation}\label{emt}
	T_{\mu\nu} = \left(\rho_m+p \right)U_{\mu}U_{\nu}+p\,g_{\nu\mu},
	\end{equation}
	where $\rho_m$ is the energy density, $p$ is the isotropic fluid pressure, and $U_{\mu}$ is the 4--velocity of the relativistic matter fluid. Also, $E_{\mu\nu}$ is the energy tensor of the electromagnetic field given by
	\begin{equation}\label{emem}
	E_{\mu\nu} = \frac{1}{4\pi}\left(F_{\mu}^{\lambda}\,F_{\nu\lambda}- \frac{1}{4}g_{\mu\nu}\,F_{\lambda\sigma}\,F^{\lambda\sigma}\right),
	\end{equation}with $F_{\mu\nu}$ representing the electromagnetic field-strength tensor given by
	\begin{equation}\label{deff}
	F_{\mu\nu} = \partial_{\mu}A_{\nu}-\partial_{\nu}A_{\mu},
	\end{equation}where $A^{\mu}=\left(\phi, \vec{A}\right)$ is the electromagnetic 4--potential. Finally, denoting $\rho_e$ to the associated density of electric charges, the 4--current density can be expressed as
	\begin{equation}\label{fcurr}
	J_{\mu} = \rho_{e} U_{\mu},
	\end{equation}
	whose temporal component $J^{0}$ is equal to the charge density $\rho_e$, and spatial components $J^{i}$ are just the usual 3--vector current components. A static spherically symmetric spacetime is described by the line element
	\begin{equation}\label{metr1}
	{\rm d}s^2= g_{\mu\nu}\,{\rm d}x^{\mu}\,{\rm d}x^{\nu},
	\end{equation}
	which can be rewritten in the form
	\begin{equation}\label{metr2}
	{\rm d}s^2= -W^2\,{\rm d}t^2+ h_{ij}\,{\rm d}x^{i}\,{\rm d}x^{j},
	\end{equation}where $W$ is the gravitational potential, so the 4--potential $A_{\mu}$ and the 4--velocity $U_{\mu}$ are defined via
	\begin{equation}\label{aest}
	A_{\mu}= -\phi \,\delta^{0}_{\mu},\qquad{\rm and}\quad
	U_{\mu}= -W \delta^{0}_{\mu}.
	\end{equation}
	Notice that we are considering only pure electric fields $A_{\mu} =\left(-\phi, 0, 0, 0\right)$, and also, $h_{ij}$ are functions of the spatial coordinates $x^i$ only with $i=1,2,3$. 
	In particular, we are interested in a class of the Guilfoyle solutions  \cite{Guilfoyle,Lemos} where the gravitational potential $W$ and the electric potential $\phi$ are related by means of the equation
	\begin{equation}\label{relwphi}
	W(\phi) =\sqrt{a \left(b-\epsilon \phi \right)^2+b_1},
	\end{equation}
	where $a$, $b$ and $b_1$ are arbitrary constant and $\epsilon = \pm 1$. Here the parameter $a$ being called the {\it Guilfoyle parameter}. As stated in \cite{Lemos(2009)}, from the set of quantities $\{\rho_m, p, \rho_e, \rho_{em}\}$ we obtain the following equation of state:
	
	\begin{equation}\label{contr1}
	p=\frac{a\,(b-\epsilon\,\phi)\,\rho_e -\epsilon\,W(\phi)\,\left[\rho_m+\left(1-a\right)\rho_{em}\right]}{3\,\epsilon\,W(\phi)},
	\end{equation}
	where $\rho_{em}$ is the electromagnetic energy density defined by means of the Weyl--Guilfoyle relation:
	
	\begin{equation}
	\rho_{em} =\frac{1}{8\pi}\frac{\left(\nabla_i \phi\right)^2}{W^2(\phi)}.
	\end{equation}

	\section{Spherical equations: general analysis}
	\label{spherical}
	The geometry of the static spherically symmetric solution for charged fluid distribution found by Guilfoyle \cite{Guilfoyle} can be written in usual Schwarzschild coordinates, $\left(t, r, \theta, \varphi \right)$ as
	\begin{equation}\label{dss}
{\rm d}s^{2}=-N(r)\,{\rm d}t^{2}+ B(r)\,{\rm d}r^2+r^{2}\left({\rm d}\theta^{2}+\sin^{2}\theta\,{\rm d}\varphi^{2}\right),
	\end{equation}
	where the structural functions $N$ and $B$ depend on the radial coordinate $r$ only, so the gauge field and 4--velocity are then given by
	
	\begin{equation}\label{gausph}
	A_{\mu}= -\phi\, \delta^{0}_{\mu},\qquad{\rm and}\qquad
	U_{\mu}= -\sqrt{N(r)}\, \delta^{0}_{\mu}.
	\end{equation}
	We now turn our attention to the stellar mass equation of the spherically symmetric solutions due to the total contribution of the matter energy density and electric energy density. Thus, the mass inside a sphere of radius $r$ can be obtained from the following relation:
	\begin{equation}\label{Mr}
	M(r) = 4\pi \int^{r}_{0} \left(\rho_m(r)+\frac{Q^2}{8\pi r^4}\right)\,r^2\,dr+\frac{Q^2}{2r},
	\end{equation}
	and therefore, the total charge enclosed in the same region becomes
	\begin{equation}\label{Q}
	Q(r) = 4\pi \int^{r}_{0}  \rho_e(r) \sqrt{B(r)}r^2 dr.
	\end{equation}
	As we have mentioned, our main goal here is to study a type of system for which the lapse function $N(r)$ connects the gravitational and electric potentials via the Majumdar--Papapetrou relation $N(r)=W^2(\phi(r)) =a \left[b-\epsilon \phi (r) \right]^2$ ($b_1=0$ in  Eq.(\ref{relwphi})), in which case one gets
	\begin{equation}
	\epsilon\, \phi(r) = b-\sqrt{\frac{N(r)}{a}}.
	\end{equation}
	Due to the additive nature of the fields, the constant $b$ can be absorbed into the potential, so without loss of generality we choose $b = 0$ and the lapse function becomes
	\begin{equation}\label{B}
	N(r)=a \, [\epsilon\, \phi(r)]^2=a \,\phi(r)^2.
	\end{equation}
   For static spherically symmetric configurations, as considered in the present research, we have written  only the components of $tt$ and $rr$ of Einstein equation (\ref{efe}), furnish the following relations \cite{Lemos, Lemos(2009)}:
	\begin{eqnarray}\label{lemos1}
	\frac{1}{r B(r)}\frac{{\rm d}}{{\rm d}r}\ln\left[N(r)\,B(r)\right]=8\pi\left(\rho_m(r)+p(r) \right),&\\\label{rA}
\frac{{\rm d}}{{\rm d}r}\left[ \frac{r}{B(r)} \right]=8\pi\left(\rho_m(r)+\frac{Q^2}{8\pi r^4} \right),&
		\end{eqnarray} whereas the first integral of the only nonzero component of Maxwell equations (\ref{fjrelation}) furnishes
	\begin{equation}\label{nzme}
	Q(r) = \frac{r^2 \phi'(r)}{\sqrt{N(r)\,B(r)}},
	\end{equation}
	where an integration constant is set to zero and a prime denotes a derivative with respect to the radial coordinate $r$. Therefore, by making use of Eq. (\ref{B}) in Eq. (\ref{nzme}), we obtain a differential form for the electric charge
	\begin{equation}
	Q(r) =-\frac{\epsilon}{2\sqrt{a}} \frac{ r^2 N^{\prime}(r)}{\sqrt{B(r)}\,N(r)}.
	\end{equation}
	Therefore, one can easily obtain an electric charge inside a spherical surface of radius $r$, if the metric functions are known. Finally, the conservation law $\nabla_\mu T^{\mu\nu}=0$ together with the Maxwell equations, lead to the hydrostatic equilibrium equation that determines the global structure of electrically charged star is obtained by requiring the conservation of mass-energy \cite{oppenvol,tolman} for the system: 
	\begin{equation}
	2p^{\prime}(r)+ \frac{N^{\prime}(r)}{N(r)}\left[\rho_m(r)+p(r) \right]- 2\frac{\phi^{\prime}(r)\, \rho_e(r)}{\sqrt{ N(r)}}=0,
	\end{equation}
which is the only non-identically zero component of the conservation equations. The first two terms on the l.h.s. comes form the gravitational force with an isotropic pressure and density, while the second term due to the Coulomb force that depends on the matter by the metric coefficient.

	\section{Guilfoyle's solutions for the interior region}
	\label{gsir}
	In this section we describe a static spherically symetric distribution of electrically charged matter in the range $0\leq r\leq r_0$. To do this, we adopt the simplest ansatz of Guilfoyle's solution \cite{Guilfoyle} given by
	
	\begin{equation}
	8\pi \rho_m+\frac{Q^2}{ r^4} = \frac{3}{R^2},
	\end{equation} where $r \leq r_0$ and $R$  is a constant and characterizes the length associated with the inverse of the total energy density. It can be also related to the parameters of the exterior solution, namely the total charge $q$ and the total mass $m$ evaluated at the junction boundary $r = r_0$,\begin{equation}
	\label{defR}\frac{1}{R^2}=\frac{2}{r_0^3}\left(m-\frac{q^2}{2 r_0}\right),
	\end{equation}where 	\begin{eqnarray}\label{Mr2}
	&m=M(r_0) = 4\pi \int^{r_0}_{0} \left(\rho_m(r)+\frac{Q^2}{8\pi r^4}\right)\,r^2\,dr+\frac{q^2}{2r_0},\\\label{qbound}
	&q=Q(r_0).
	\end{eqnarray}
	
	 Using this additional assumption, Guilfoyle \cite{Guilfoyle} found the exact solutions of Einstein--Maxwell equations, which can be summarized as follows: the structural functions become
	\begin{equation}\label{Bg}
	B(r) = 1-\frac{r^2}{R^2},\end{equation}\begin{equation}\label{Ng}
	N(r) = \left[\frac{(2-a)}{a}\,F(r)\right]^\frac{2a}{(a-2)},
	\end{equation}
	with the function $F(r)$ defined as
	\begin{equation}\label{Fg}
	F(r) = c_o \sqrt{1-\frac{r^2}{R^2}}-c_1,
	\end{equation}
	where the integration constants $c_0$ and $c_1$, obtained via the junction conditions, are given by
	\begin{eqnarray}\label{c0}
	&c_o= \frac{R^2}{r^2_0} \left(\frac{m}{r_0}-\frac{q^2}{r^2_0}\right)\left(1-\frac{r^2_0}{R^2}\right)^{-1/a},\\\label{c1}
	&c_1= c _o \sqrt{1-\frac{r^2_0}{R^2}}
	\left[1-\frac{a}{(2-a)}\frac{r^2_0}{R^2}\left(\frac{m}{r_0}-\frac{q^2}{r^2_0}\right)^{-1} \right].
	\end{eqnarray}
	Thus, from Eq. (\ref{B}), the electric potential is given by
	\begin{equation}\label{potelg}
	\phi(r) = \frac{\epsilon}{\sqrt{a}}\left[\frac{(2-a)}{a}\,F(r)\right]^\frac{a}{(a-2)},
	\end{equation}
	whereas the fluid quantities are given by
	\begin{equation}\label{rhomg}
	8\pi \rho_m(r) =\frac{3}{R^2}\left[1-\frac{a\,c^2_0}{3R^2\,(2-a)^2}\frac{r^2}{F^2(r)}\right],
	\end{equation}
	\begin{equation}
	\begin{split}\label{pg}
	8\pi p(r) =-\frac{1}{R^2}&\left[1+\frac{a\,c^2_0}{R^2\,(2-a)^2}\frac{ r^2}{F^2(r)}+\right.\\
	&\left.-\frac{2a}{(2-a)} \frac{F(r)+c_1}{F(r)}\right],
	\end{split}
	\end{equation}
	
	\begin{equation}
	8\pi \rho_e(r) =\frac{\epsilon \sqrt{a}}{R^4(2-a)}\frac{ r^2}{F^2(r)}\left[c^2_0+\frac{3F(r)\,(F(r)+c_1)}{r^2}\right].
	\end{equation} Finally, inserting the above relations into Eqs.(\ref{Mr}) and (\ref{Q}), the mass and electric charge functions $M(r)$ and $Q(r)$ are found to be
	\begin{equation}
	M(r) =\frac{r^3}{2R^2}\left[1+\frac{a\,c_0^2}{R^2(2-a)^2}\frac{r^2}{ F^2(r)}\right],
	\end{equation}and
	\begin{equation}
	Q(r) =\frac{\epsilon \sqrt{a}\,c_0}{R^2 \,(2-a)}\frac{ r^3}{F(r)},
	\end{equation}respectively.
	
	\section{review of gravastar models}
	\label{rgm}
	This section presents a quick review of the main features of two Gravastar models, which possess some interesting properties for our subsequent study.   
	\subsection{Mazur--Mottola model}
	The first model presented here is the so--called  {\it Mazur \& Mottola gravastar model} \cite{Mazur}. According to the model, the interior of the gravastar is a de Sitter spacetime surrounded by a layer of ultra--stiff matter, while the exterior is then suitably matched by a Schwarzschild spacetime i.e., there are five different regions (including two thin shells), each with own features:
	\begin{enumerate}
		\item[(i)] Inside the gravastar $0 \leq r <  r_1$, a de Sitter spacetime with $p=-\rho$.
		\item[(ii)] An interior thin shell at $r_1\lesssim r_s$ with surface density $\sigma_-$ and surface tension $\vartheta_-$.
		\item[(iii)] A finite layer of ultra--stiff matter, $p=\rho$, placed at $r_1< r < r_2$.
		\item[(iv)] An exterior thin shell at $r_2\gtrsim r_s$ with surface density $\sigma_+$ and surface tension $\vartheta_+$.
		\item[(v)] An exterior Schwarzschild vacuum $r>r_2$ with $p =\rho= 0$.
	\end{enumerate}

	These features are the most important for a gravastar model having negative central pressure, positive density and the absence of event and cosmological horizons.  Notice that here $\rho$ is the energy density and $p$ is the isotropic pressure of the gravastar, whereas the interior has a constant energy density given by $\rho_{\text{int}} =3 H_0^2/8\pi \geq 0$.
	From a physical point of view, region (iii) is the most important one because there is where the non--trivial model for the gravastar can be specified.

\subsection{The thin--shell gravastar model}
The second model to be reviewed is the {\it Visser \& Wiltshire gravastar model} \cite {Visser}, in which the authors tried to determine the possibility of dynamically testing the stability of the gravastar model against radial perturbations. In this sense, the scenario allows a precise formation mechanism, in which the number of layers of the original model is reduced from five to three. Therefore, the most important features of the model are:
\begin{enumerate}
\item[(i)] Inside the gravastar, $r < r_0$, a de Sitter spacetime with $p=-\rho$  is assumed, together with the strong condition $\rho>0$.
\item[(ii)] The spacetime is assumed to be free of singularities everywhere. In order to avoid both event and cosmological horizons, a single thin shell with a surface density $\sigma$ and surface tension $\vartheta$ is placed at $r=a\gtrsim r_s$.
\item[(iii)] The interior spacetime is matched to the exterior vacuum at the junction interface $\Sigma$ situated at $r=a$.
\end{enumerate}

	A particular illustration in this model that offers a de Sitter interior solution was matched smoothly to the exterior Schwarzschild geometry at a junction surface, composed of a thin shell with surface energy density $\sigma$ and surface pressure $\mathcal{P}$. Using this technique, we obtain a condition for dynamical stability for the thin shell against radial perturbations in terms of a thin shell's equation of motion given by $ \frac{1}{2}\dot{a^2} + V(a) = 0 $, with R $\equiv$ $\frac {dR}{d\tau}$ and $\tau$ being the proper time of the timelike hypersurface. In this context, we perform a similar procedure outlined by Visser \& Wiltshire and analyze the stability of gravastars against radial perturbations.

	\section{ Construction of gravastar model}
	\label{cgm}
This section is devoted to model a specific gravastar geometry by matching an interior de Sitter spacetime with an exterior Reissner--Nordstr\"om spacetime at junction interface $\Sigma$. So, the structural functions $N(r)$ and $B(r)$ are given by
	\begin{equation}\label{Brr}
	N(r) = B(r)^{-1} =
	\begin{cases}
	1-\frac{r^2}{R^2} & \text{when } r < a(\tau),\\
	\\
	1-\frac{2m}{r}+\frac{q^2}{r^2} & \text{when}\, r > a(\tau),
	\end{cases}
	\end{equation}
	where $r= a(\tau )$ is the timelike hypersurface at which the infinitely thin shell is located at the proper time $\tau$.
	Now we focus on the analysis of the separating surface between the two spacetimes, which is defined by the radial coordinate $r_0=a(\tau)$. So, in the procedure of matching the interior de Sitter spacetime to the exterior Reissner--Nordstr\"om spacetime, we must consider two different manifolds: an exterior $\mathcal{V^{+}}$ and an interior $\mathcal{V^{-}}$, which are joined at the surface layer $\Sigma$, and induce the metrics $g^{+}_{ij}$ and $g^{-}_{ij}$, respectively. Thus, a single manifold $\mathcal{V} = \mathcal{V^{+}}\cup \mathcal{V^{-}}$ is obtained by gluing them together at their boundaries. The induced metric on the hypersurface is a timelike hypersurface $\Sigma$, defined by a parametric equation in the form $f\left(x^\mu(\xi^i)\right)=0$, where $\xi ^i =\left(\tau, \theta, \varphi\right)$ denotes the intrinsic coordinates on $\Sigma$. In order to describe the position of the junction surface, we consider $\xi^{\mu} (\tau, \theta, \varphi)$ = $(t(\tau), a(\tau), \theta, \varphi)$, and the induced metric on the hypersurface can be written as
	\begin{equation}\label{induced1}
	{\rm d}s^2 = -{\rm d}\tau^2 + a^{2}(\tau)\,{\rm d}\Omega^2_{2},
	\end{equation}
	where $\tau$ is the proper time along the hypersurface $\Sigma$.
	
Next, we shall use the Darmois--Israel formalism to determine the relation between the geometry and thin layer of matter at the shell across a junction surface, which is given by the Lanczos equations \cite{Israel,Musgrave}:
	\begin{equation}\label{Sij}
	S^{i}_{j} = -\frac{1}{8\pi}\left(\kappa^i_j-\delta^i_j\,\kappa^m_m\right),
	\end{equation}
where $S_{ij} $ represents the surface energy--momentum tensor and the discontinuity in the second fundamental form or extrinsic curvatures across a junction surface is given by the quantity $\kappa_{ij} = K^{+}_{ij}-K^{-}_{ij}$, which is associated with both sides of the shell. Also, the extrinsic curvature $K^{\pm}_{ij}$ is defined on each side of the shell, and is given by
	\begin{equation}
	K^{\pm}_{ij} = -\eta^{\pm}_{\nu}\left(\frac{\partial^2 x^{\nu}}{\partial\xi^i \partial\xi^j}+\Gamma^{\nu\pm}_{\alpha\beta}\frac{\partial x^{\alpha}}{\partial\xi^{i}}\frac{\partial x^{\beta}}{\partial\xi^{j}}\right),\label{9}
	\end{equation}
where $\eta_{\nu}$ represents the unit normal vector to $\Sigma$ and $\xi^i$ represents the intrinsic coordinates.
Thus, at the hypersurface $\Sigma$, whose parametric equation is given by $f\left(x^\mu(\xi^i)\right)=0$, the respective unit 4--normal vectors  $\eta^{\pm}_{\nu}$ to $\Sigma$ are
	\begin{equation}
	n_{\mu}^{\pm} = \pm\left| g^{\alpha\beta}\frac{\partial f}{\partial x^{\alpha}}\frac{\partial f}{\partial x^{\beta}}\right| ^{-\frac{1}{2}}\frac{\partial f}{\partial x^{\mu}}, \label{10}
	\end{equation}
where the unitary conditions $n_{\mu}n^{\mu}= +1$ and $n_{\mu}e^{\mu}_{(i)} = 0$ are oriented outwards from the origin. By using Eq. (\ref{10}), normal vectors may be determined from the interior and exterior spacetimes given in Eq. (\ref{Brr}), so they become
	\begin{eqnarray}\label{normal1} &n_{-}^{\mu}=\left(\frac{\dot{a}}{\left(1-\frac{a^{2}}{R^{2}}\right)},
	\sqrt{\left(1-\frac{a^{2}}{R^{2}}\right)+\dot{a}^{2}}, 0, 0\right),&\\ \label{normal}
	&n_{+}^{\mu}=\left(\frac{\dot{a}}{1-\frac{2M}{a}+\frac{q^{2}}{a^{2}}} ,\sqrt{1-\frac{2M}{a}+\frac{q^{2}}{a^{2}}+\dot{a}^{2}}, 0, 0\right),&
	\end{eqnarray}
	where the ($\pm$) superscripts correspond to the exterior and
	interior spacetimes, respectively.
	
	It is well known that the discontinuity of the extrinsic
	curvature $\kappa_{ij}$ can be written in a simple form due to spherical symmetry as
	$\kappa^i_{j}= \text{diag}\left(\kappa^{\tau}_{\tau},\kappa^{\theta}_{\theta},\kappa^{\theta}_{\theta}\right)$ with $\kappa^{\theta}_{\theta}=\kappa^{\varphi}_{\varphi}$.
	By employing the expressions through (\ref{Sij}), we find the non--vanishing components of surface stress--energy tensor that can be written in terms of
	$S^i_j = \text{diag}\left(-\sigma, \cal P, \cal P \right)$, where $\sigma$ is
	the surface energy density and $P$ is the surface pressure.
	
	Now, using Eq. (\ref{Brr}) and Eq. (\ref{9}) the non--trivial components of the extrinsic curvature
	are given by
	\begin{eqnarray}
	K_{\;\;\theta}^{\theta\;-} & = & \frac{1}{a}\,\sqrt{\left(1-\frac{a^{2}}{R^{2}}\right)+\dot{a}^{2}}\;,\label{genKplustheta1}\\
	K_{\;\;\tau}^{\tau\;-} & = & \frac{\left(\ddot{a}-\frac{a}{R^{2}}\right)}{\sqrt{\left(1-\frac{a^{2}}{R^{2}}\right)+\dot{a}^{2}}},\label{genKminustautau1}
	\end{eqnarray}
	\begin{eqnarray}
	K_{\;\;\theta}^{\theta\;+} & = & \frac{1}{a}\,\sqrt{1-\frac{2M}{a}+\frac{q^{2}}{a^{2}}+\dot{a}^{2}}\;,\label{genKplustheta}\\
	K_{\;\;\tau}^{\tau\;+} & = & \frac{\ddot{a}+\frac{M}{a^2}-\frac{q^{2}}{a^{3}}}
	{\sqrt{1-\frac{2M}{a}+\frac{q^{2}}{a^{2}}+\dot{a}^{2}}} \,,\label{genKminustautau}
	\end{eqnarray}
	where dot denotes a derivative with respect to $\tau$.
	Therefore, using Eqs. (\ref{genKplustheta1}, \ref{genKminustautau1}, \ref{genKplustheta}, \ref{genKminustautau}) into the Lanczos equations (\ref{Sij}), we obtain that the energy density $\sigma \equiv -\kappa^{\theta}_{\theta}/4\pi$ and the pressure at the junction surface ${\cal{P}} \equiv (\kappa^{\tau}_{\tau}+\kappa^{\theta}_{\theta})/8\pi$
	becomes
	\begin{equation}\label{sigm1}
	\sigma = -\frac{1}{4\pi a}\left[\sqrt{1-\frac{2M}{a}+\frac{q^{2}}{a^{2}}+\dot{a}^{2}}
	-\sqrt{1-\frac{a^{2}}{R^{2}}+\dot{a}^{2}}\;\right],
	\end{equation}
	and
	\begin{equation}\label{pres1}
	{\cal{P}} = \frac{1}{8\pi a}\left[\frac{1-\frac{M}{a}+\dot{a}^{2}+a\ddot{a}}{\sqrt{1-\frac{2M}{a}+\frac{q^{2}}{a^{2}}+\dot{a}^{2}}}-\frac{1-\frac{2a^2}{R^{2}}+\dot{a}^{2}+a\ddot{a}}{\sqrt{1-\frac{a^{2}}{R^{2}}+\dot{a}^{2}}}\right],
	\end{equation} respectively.
Notice that the surface density $\sigma$ has the opposite sign to that of the surface pressure ${\cal P}$, and also that the all energy--momentum that plunges into the thin shell still satisfies an energy conservation law,	$\nabla_i S^{ij} = 0$ by virtue of $\nabla^i \left(\kappa^{ij}-\delta _{ij}\kappa\right) = 0$	at the junction interface.
	
	Now, using two equations, (\ref{sigm1}) and (\ref{pres1}), it is easy to check the energy conservation equation is fulfilled:
	\begin{equation}\label{conserv}
	\frac{\textrm{d}}{\textrm{d}\tau}\left(\sigma a^2\right)+{\cal P} \frac{\textrm{d}}{\textrm{d}\tau}\left( a^2 \right)=0\,,
	\end{equation}
	which immediately follows
	\begin{equation}
	\dot{\sigma}= -2\left(\sigma + {\cal P}\right)\frac{\dot{a}}{a}.
	\end{equation}
	Integrating out the above equation, it yields
	\begin{equation}\label{sigprim}
	\sigma^{\prime} = -\frac{2}{a}\left(\sigma + {\cal P}\right),
	\end{equation}
	where primes and dots denote differentiation with respect to $a$ and $\tau$, respectively.
	The first term on the left side of expression (\ref{conserv}) represents the internal energy change of the shell, while the work done by internal forces of the shell is given in the second term.
	
	Now, by taking into account the Eqs. (\ref{sigm1}-\ref{pres1}) and substituting into Eq. (\ref{sigprim}), we obtain the following expression 
	\begin{equation}
	\sigma^{\prime}  = \frac{1}{4\pi a^2}\left[\frac{1-\frac{3M}{a}+\frac{2q^2}{a^2}+\dot{a}^{2}-a\ddot{a}}{\sqrt{1-\frac{2M}{a}
			+\frac{q^{2}}{a^{2}}+\dot{a}^{2}}}-\frac{1+\dot{a}^{2}-a\ddot{a}}
	{\sqrt{1-\frac{a^{2}}{R^{2}}+\dot{a}^{2}}}\right],
	\end{equation}
	which plays an important role in determining the stability regions when
	the static solution $a_0$ is being considered.
	
	According to Ref. \cite{Poisson,Lobo(2006)}, the total surface mass of the thin shell is given by $m_s= 4\pi a^2\sigma $. For this case the total mass of the system $M$ evaluated at a static solution $a_0$, is given by (by rearranging the Eq.(\ref{sigm1}))
	\begin{equation}
	M= \frac{a_0^3}{2R^2}+\frac{q^2}{2a_0}+m_s\left[\sqrt{1-\frac{a_0^{2}}{R^{2}}}-\frac{m_s(a_0)}{2a_0}\right].
	\end{equation}
	Note that $M(a)$ is the total active gravitational mass and $q(a)^{2}/2a$ is the mass equivalent to the electromagnetic field.
	
	\section{Junction conditions}\label{jc}
	Keep in mind that a gravastar model does not possess an event horizon.
	For instance, if the thin--shell transition layer $\Sigma$ is located at $r = a(\tau)$, then to avoid horizon formation we demand
	\begin{equation}
	0< \Bigl\rvert \frac{r}{R}\Bigl\rvert  < 1 ~~ \text{and}~~0< \Bigl\rvert\frac{2M}{r}-\frac{q^2}{r^2}\Bigl\rvert < 1.
	\end{equation}
	For the present analysis, one can obtain the solution of the non--rotating thin shell gravastar when the spacetimes given by the metrics (\ref{Brr}) are matched at $a$. The restriction of charge--to--mass ratio $|q|/M < 1$ for the Reissner-Nordstr\"om spacetime corresponds to two horizons, namely the Cauchy and event horizons $r_{\pm}= M\left(1\pm \sqrt{1- q^2/M^2}\right)$, which when $|q|/M \rightarrow 1$, they are glued into a single horizon. For the case when $|q|/M > 1$,  it is a naked singularity. Moreover, we consider the case when $|q|/M \leq 1$ in order to avoid the horizon from geometry, where	$ a> r_+ = M\left(1+ \sqrt{1- q^2/M^2}\right)$. We develop the rest of the section	by assuming $|q|/M \leq 1$, where the junction surface $r = a$ is situated outside the event horizon.

	\section{Static gravastars}\label{sg}
	Now, we resolve the static case, which is given by taking into account $\dot{a}= \ddot{a} =0$. In this case, Eqs. (\ref{sigm1}) and (\ref{pres1}) reduce to 
	\begin{equation}\label{sigstat}
	\sigma (a_0)  =   -\frac{1}{4\pi a_0}\left[\sqrt{1-\frac{2M}{a_0}+\frac{q^{2}}{a^{2}_0}}
	-\sqrt{1-\frac{a^{2}_0}{R^{2}}}\;\right],
	\end{equation}
	\begin{equation}\label{prestat}
	{\cal P}(a_0) = \frac{1}{8\pi a_0}\left[\frac{1-\frac{M}{a_0}}{\sqrt{1-\frac{2M}{a_0}
			+\frac{q^{2}}{a^{2}_0}}}-\frac{1-\frac{2a^2_0}{R^{2}}}
	{\sqrt{1-\frac{a^{2}_0}{R^{2}}}}\right].
	\end{equation}
	It is worth noting that the gravastar solution is also considered a self-gravitating object  which can avoid the formation of the black hole horizon. Hence, a gravastar model can be considered one type of compact objects where the surface redshift is an important source of information. We expect that the redshifts of the gravastar are higher than any ordinary objects. The surface gravitational redshift is defined by $z= \Delta \lambda/\lambda_{e} $ = $\lambda_{0}/\lambda_{e}$, where $\Delta$ is the fractional change between the observed wavelength $\lambda_{0}$ and the emitted wavelength $\lambda_{e}$.
	
	Thus, according to our notation, the redshift factor ($z_{a_0}$) related to the
	\begin{equation}
	z_{a_0}= -1+\Bigl\rvert g_{tt}(a_0)\Bigl\rvert ^{-1/2},
	\end{equation}
	Now, the behavior of the surface redshift is not larger than $z = 2$ \cite{Buchdahl} for a static perfect fluid sphere. This value may increase up to 3.84, when we consider anisotropic fluid spheres \cite{Ivanov}.
    
    \begin{figure}[h!]
		\includegraphics[width=0.36\textwidth]{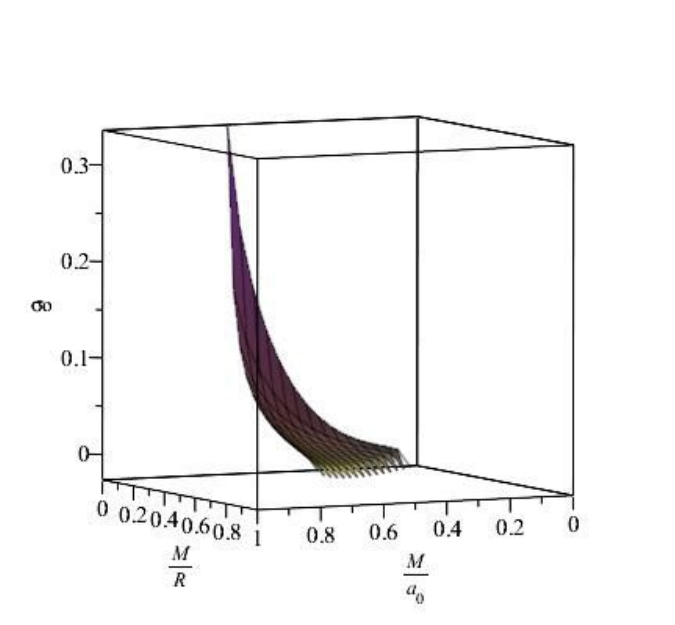} %
		\caption{\small \textit{Unification diagram for surface energy density $\sigma (a_0)$.	We have considered the dimensionless parameters, $x = M/a_0$, $y = M/R$. The qualitative values adjacent for drawing the graphs are $w =0.5$ and $R = 3$.		Note that surface energy density and surface pressure both are positive in the range. See the appendix for more details. } }\label{fig1}
	\end{figure}
	
	In order to specify an equilibrium solution on the thin--shell gravastar, we introduce the dimensionless configuration varibles defined by \cite{Moruno} \begin{equation}
	\label{advar} x\equiv \frac{M}{a_0},\qquad y\equiv \frac{M}{R},\quad {\rm and}\qquad w\equiv \frac{q}{M}.
	\end{equation}  Therefore, assuming the condition of $|q| < a_0 < R$, the surface energy density $\sigma(a_0)=\sigma(x)$ and the surface pressure $\mathcal{P}(a_0)=\mathcal{P}(x)$ can be written as
	\begin{equation}
	\sigma (x) = -\frac{1}{4\pi R}\,\frac{x}{y}\left(\sqrt{1-2x+w^{2}\,x^2}
	-\sqrt{1-\frac{y^2}{x^{2}}}\;\right),
	\end{equation}
	and
	\begin{equation}
	\mathcal{P}(x)  = \frac{1}{8\pi R}\frac{x}{y}
	\left(\frac{1-x}{\sqrt{1-2x+w^2\,x^2}}-\frac{1-2\frac{y^2}{x^{2}}}
	{\sqrt{1-\frac{y^{2}}{x^{2}}}}\right).
	\end{equation}
	\begin{figure}[h!]
		\includegraphics[width=0.35\textwidth]{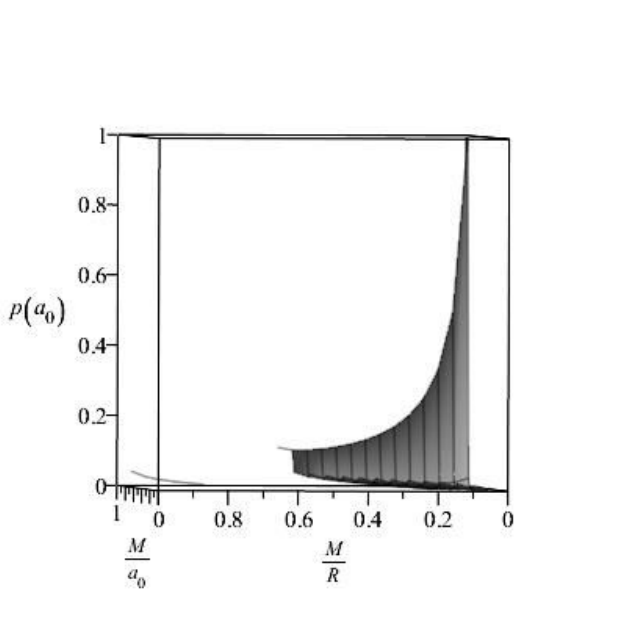} %
		\caption{\small \textit{In this plot we show  surface pressure. The values of parameters for graphical representation are $w =0.5$ and $R = 3$.} }\label{fig2}
	\end{figure}
	
	\section{LINEARIZED STABILITY OF GRAVASTARS}\label{lsog}
	In this section we study the stability of the gravastar model about the static solutions $a_0$, also known as the linearized stability of the solutions.
	In order to test whether the equilibrium solution is stable or not, we rearrange expression (\ref{sigm1}) in the following suggestive form
	\begin{equation}
	\frac{1}{2}\dot{a}^2+V(a) =0,
	\end{equation}
	which is known as the thin--shell equation of motion and the effective potential
	is given by
	\begin{equation}
	V(a)=-\frac{256 \sigma^4(a) \pi^4 a^4 -32 \pi^2 a^2\, \sigma^2(a)\, \Xi(a)+\zeta^2(a)}{64\, \sigma^2(a) \pi^2 a^2},
	\end{equation}
	where, for computational convenience, we introduce the functions $\Xi(a)$ and $\zeta(a)$ by means the following definitions
	\begin{eqnarray}
	\Xi(a)&=&\left(1-\frac{2M}{a}+\frac{q^2}{a^2}\right)+\left(1-\frac{a^2}{R^2}\right),\\
	\zeta(a)&=&\left(1-\frac{2M}{a}+\frac{q^2}{a^2}\right)-\left(1-\frac{a^2}{R^2}\right).
	\end{eqnarray}
	This allows us to write the potential in a second-order differential form as
	\begin{widetext}
		\begin{equation}\label{vpp}
		V^{\prime\prime}(a)=\frac{1}{32 \sigma^4 a^4 \pi^2}\Big[ \Delta_1  \sigma''+16 \sigma^4 \pi^2 \Xi'' a^4-\sigma^2 \zeta a^2 \zeta''+\Delta_2 \sigma'+\Delta_3(\sigma')^2-256 \sigma^6 \pi^4 a^4 -\sigma^2 \,\zeta'^2 a^2+4 \sigma^2 \zeta' \zeta a-3 \sigma^2 \zeta^2\Big],
		\end{equation}
	\end{widetext}
		where
		\begin{eqnarray}\notag
		\Delta_1 &=& -256 a^6 \pi^4 \sigma^5+a^2 \zeta^2 \sigma,\\\notag
		\Delta_2 & =& -1024 a^5 \pi^4\sigma^5+4a^2 \sigma \zeta \zeta'-4a \zeta^2 \sigma,\\\notag
		\Delta_3 &=& -256 a^6 \pi^4 \sigma^4-3 a^2 \zeta^2.
		\end{eqnarray}
		
		In the study of the stability of a static solution at $a_0$, we perform a Taylor expansion of $V(a)$ about $a_0$ until second--order terms, given by
		\begin{eqnarray}\nonumber
		V(a) &=&  V(a_0) + V^\prime(a_0) ( a - a_0) +
		\frac{1}{2} V^{\prime\prime}(a_0) ( a - a_0)^2 +\\  &&+O\left[( a - a_0)^3\right],
		\end{eqnarray}
		where a prime denotes derivative with respect to $a$.
		Here we adapt and apply the criteria for stability analysis of a static configuration 
		at $a = a_0$, which requires $V(a_0)= V'(a_0)=0$, and the condition for stability is that  $V^{\prime\prime}(a_0) > 0$ to guarantee that the second derivative of the potential is positive.
		In order to determine the stability of the gravastar we first calculate  $V^{\prime\prime}(a_0)$. 
		Using Eq. (\ref{sigprim}) and by introducing a new parameter $\eta_0={\cal P}'(a_0)/\sigma'(a_0)$ into Eq. (\ref{vpp}), we obtain
		\begin{widetext}
			\begin{eqnarray}\nonumber
					V^{\prime\prime}(a_0)& = &\frac{1}{32\,\pi^2 \sigma_0^4 a_0^8 R^4}\left\{ -1024 \left(\eta_0+\frac{3}{4}\right)\pi^4 R^4 a_0^8 \sigma_0^6-1024  \sigma_0^5 \pi^4 R^4a_0^8 \left(\eta_0+\frac{3}{2}\right)p_0\,+\right.\\\nonumber
					&-&64\, \left[  \left( 16\,a_0^{4}p_0^{2}\pi^{2}+Ma_0-\frac{3}{2}\,{q}^{2}
					\right) R^{2}+\frac{1}{2}\,a_0^{4} \right] \pi^{2}R^{2}a_0^{4} \sigma_0^{4}-48\,p^{2} \left[  \left( Ma_0-\frac{1}{2}\,{q}^{2} \right) {R}^{2}-\frac{a_0^4}{2} \right] ^{2}+\\\notag
					&+& 16\,p \left[  \left( M \left( \eta_0-\frac{1}{2} \right) a_0-\frac{1}{2}\,{q}^{2} \left(
					\eta_0+\frac{3}{2} \right)  \right) {R}^{2}-1/2\, \left( \eta_0-\frac{13}{2} \right) a_0^{4} \right]  \left[\left( Ma_0-\frac{q^2}{2} \right) {R}^{2}-\frac{a_0^4}{2} \right] \sigma_0+\\\notag
					&+& \left[  \left( 16\,{M}^{2}a_0^{2}\eta_0-16\,M{q}^{2} \left( \eta_0-\frac{1}{4}
					\right) a_0+4\, \left( \eta_0-\frac{3}{4} \right) {q}^{4} \right) {R}^{4}-16\,
					\left( M \left( \eta_0-\frac{3}{4} \right) a_0-\frac{1}{2}\,{q}^{2} \left( \eta_0-\frac{1}{4}
					\right)  \right) a_0^{4}{R}^{2}\right.\\
					&+& \left. \left. 4\, \left( \eta_0-{\frac {15}{4}}
					\right) a_0^{8} \right]  \sigma_0^{2}
					\right\}.
			\end{eqnarray}
		\end{widetext}
Therefore, by demanding that $V^{\prime\prime}(a_0)$ $>$ 0, we find an inequality for the $\eta_0$ parameter, which yields to
\begin{widetext}
		\begin{eqnarray}\label{eta1}
		\eta_0>\frac{768\, \sigma_0^6 {\pi }^{4}{R}^{4}a_0^8+1536\, \sigma_0^5{\pi }^{4}{R}^
			{4}a_0^8 p_0
			+64 \pi^2 R^2 a_0^4 \sigma_0^4 \Theta_1+\sigma^2_0 \Theta_2+8\pi \sigma a_0 \Theta_3+48 p_0^2 \Theta_4}{4\,\sigma_0  \left( \sigma_0 +p_0 \right) 
			\left( -256\,  \sigma_0^4 {\pi }^{4}a_0^8 {R}^{4}+4\,{M}^{2}a_0^2{R}^{4}-4\,Ma_0{R}^{4}{q}^{2}-4\,Ma_0^5
			{R}^{2}+{R}^{4}{q}^{4}+2\,a_0^4{R}^{2}{q}^{2}+a_0^8\right) 
		},
		\end{eqnarray}		
		\[ \text{if},~~~~~~
		\sigma_0  \left( \sigma_0 +p_0 \right) 
		\left( -256\,  \sigma_0^4 {\pi }^{4}a_0^8 {R}^{4}+4\,{M}^{2}a_0^2{R}^{4}-4\,Ma_0{R}^{4}{q}^{2}-4\,Ma_0^5
		{R}^{2}+{R}^{4}{q}^{4}+2\,a_0^4{R}^{2}{q}^{2}+a_0^8\right)>0,
		\]
		and 
		\begin{eqnarray}\label{eta2}
		\eta_0<\frac{768\, \sigma_0^6 {\pi }^{4}{R}^{4}a_0^8+1536\, \sigma_0^5{\pi }^{4}{R}^
			{4}a_0^8 p_0
			+64 \pi^2 R^2 a_0^4 \sigma_0^4 \Theta_1+\sigma^2_0 \Theta_2+8\pi \sigma a_0 \Theta_3+48 p_0^2 \Theta_4}{4\,\sigma_0  \left( \sigma_0 +p_0 \right) 
			\left( -256\,  \sigma_0^4 {\pi }^{4}a_0^8 {R}^{4}+4\,{M}^{2}a_0^2{R}^{4}-4\,Ma_0{R}^{4}{q}^{2}-4\,Ma_0^5
			{R}^{2}+{R}^{4}{q}^{4}+2\,a_0^4{R}^{2}{q}^{2}+a_0^8\right) },
		\end{eqnarray}
		\[ \text{if},~~~~~~
		\sigma_0  \left( \sigma_0 +p_0 \right) 
		\left( -256\,  \sigma_0^4 {\pi }^{4}a_0^8 {R}^{4}+4\,{M}^{2}a_0^2{R}^{4}-4\,Ma_0{R}^{4}{q}^{2}-4\,Ma_0^5
		{R}^{2}+{R}^{4}{q}^{4}+2\,a_0^4{R}^{2}{q}^{2}+a_0^8\right)<0. \]
		For notational simplicity, in Eqs. (\ref{eta1}) and (\ref{eta2}), we use
		\begin{eqnarray}\notag
		\Theta_1 &=&  \left( 16\,a_0^4\, p_0^2\, \pi^{2}+M\,a_0-\frac{3\,q^2}{2} \right) {R}^{2}+\frac{a_0^4}{2},\\\notag
		\Theta_2 &=& \left( -4\,Ma_0{q}^{2}+3\,{q}^{4} \right) {R}^{4}+2a_0^4 \left(q^2-6\,Ma_0\right) {R}^{2}+15\,a_0^8,\\\notag
		\Theta_3 &=& \left[  \left( Ma_0+\frac{3\,q^2}{2} \right) {R}^{2}-\frac{13\,a_0^4}{2} \right] 
		\left[\left( M\,a_0-\frac{q^2}{2} \right) {R}^{2}-\frac{a_0^4}{2} \right],\\ \notag
		\Theta_4 &=& \left[\left( Ma_0-\frac{q^2}{2} \right) {R}^{2}-\frac{a_0^4}{2} \right] ^
		{2}.
		\end{eqnarray}
\end{widetext}	
	
	\begin{figure}[h!]
		\includegraphics[width=0.35\textwidth]{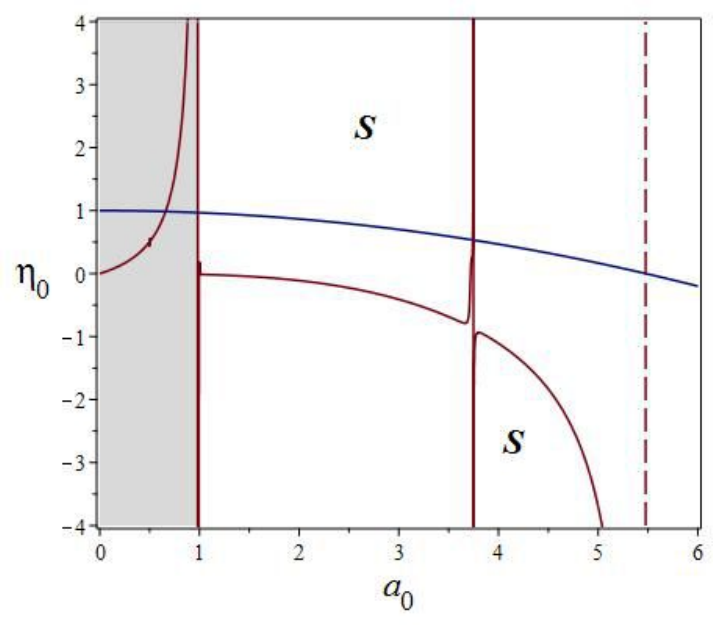} %
		\caption{\small \textit{In this plot we show stability regions in terms of $\eta_0$ as a function of  $a_{0}$ for $M=1$, $q=1$. The shaded region is for $a_0<r_h$, and the dashed vertical line is for $a_0=R$.} }\label{fig3}
	\end{figure}

	\begin{figure}[h!]
\includegraphics[width=0.35\textwidth]{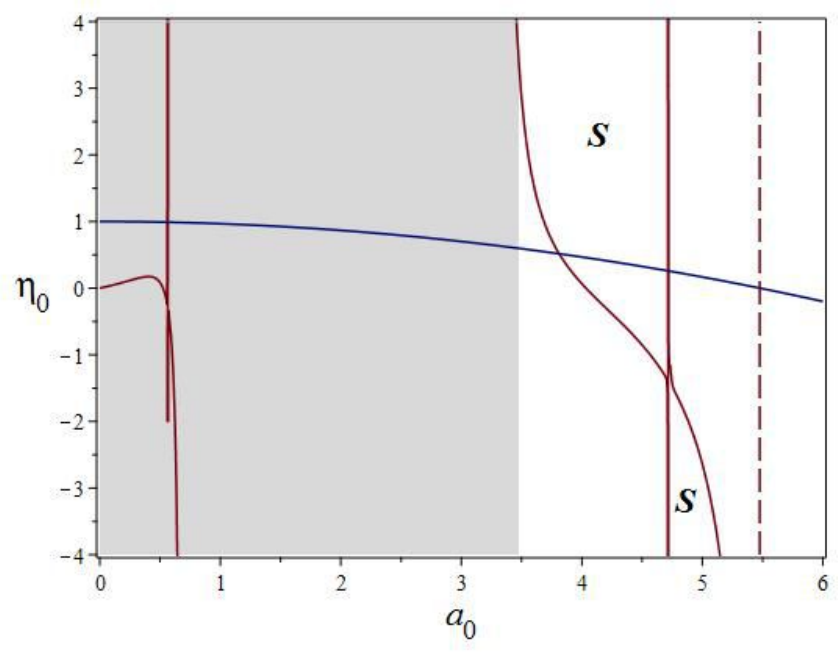} %
	\caption{\small \textit{Stability regions in terms of $\eta_0$ as a function of  $a_{0}$ for $M=2$, $q=1.5$. The shaded region is for $a_0<r_h$ and the dashed vertical line is for $a_0=R$. } }\label{fig4}
\end{figure}
	
	\begin{figure}[h!]
		\includegraphics[width=0.35\textwidth]{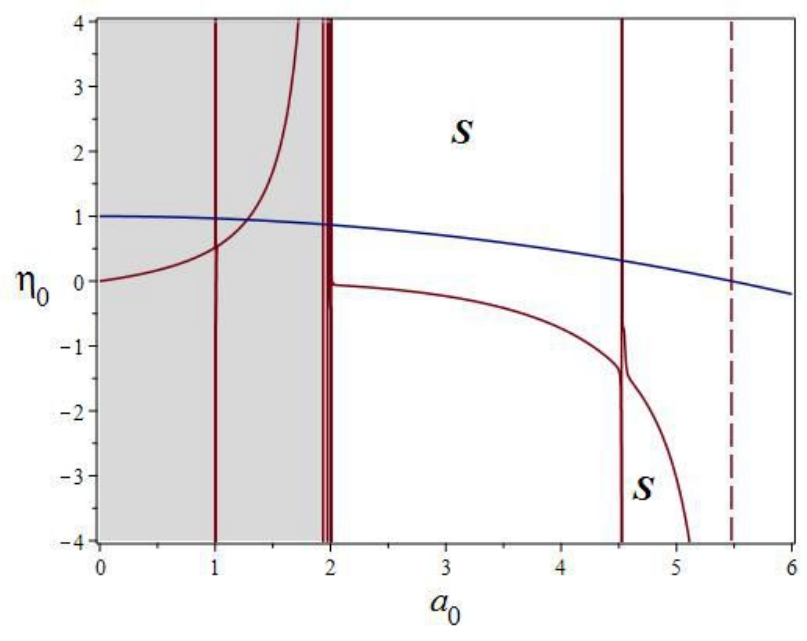} %
		\caption{\small \textit {Stability regions in terms of $\eta_0$ as a function of  $a_{0}$ for $M=2$, $q=2$. The shaded region is for $a_0<r_h$ and the dashed vertical line is for $a_0=R$.} }\label{fig5}
	\end{figure}

	Obviously, to ensure the stability, we demand that in equilibrium configurations satisfy the usual condition  $V^{\prime\prime}(a_0) > 0$. To justify our assumption for a given set of parameters and find the range of $a_0$ for which $V^{\prime\prime}(a_0) > 0$, we use a graphical representation due to the complexity of the expression $V^{\prime\prime}(a_0)$. In Figs. (\ref{fig3},\ref{fig4},\ref{fig5}), the stable solutions are displayed for different values of $q, M$ and $R$. From Eqs. (\ref{eta1}-\ref{eta2}), we find that the regions in a plot $\eta_{0} \times a_0$ where the stability conditions are satisfied.
	
	To illustrate of the above stability conditions, we use $\eta_0$ as a parameter so that there 
	is no need to specify any surface equation of state. The parameter $\eta_0$ is normally interpreted
	as the speed of sound, which should lie within the limit (0, 1] based on the physical requirement.
	However, $\eta_0$ may lie outside the range of (0, 1] on the surface layer and for an extensive discussion
	see Refs.\cite{Poisson,Lobo(2013)}.  
	\section{Stability analysis by using the surface mass of the thin shell}
	\label{sts}
	
	In this section we  study the gravastar stability through the surface mass of the thin shell, following \cite{Moruno}, which is given by $m_s = 4 \pi a^2\sigma$. For the stability analysis, we do not need to introduce a particular parameter; rather, we can simply choose $\sigma(a)$, or equivalently $m_s$, as an
	arbitrarily specifiable function that encodes the whole gravastar stability.
	
	For our purpose in the exterior Reissner-Nordstr$\ddot{\text{o}}$m spacetime, with de Sitter interior geometry,
	the surface mass of the thin shell for static configuration is given by
	\begin{equation}
	m_s(a_0) =  -a_0 \left[\sqrt{1-\frac{2M}{a_0}+\frac{q^{2}}{a_0^{2}}}-\sqrt{1-\frac{a^{2}_0}{R^{2}}}\;\right],
	\end{equation}
	while for first--order differentiation we have
	\begin{equation}
	m_s^{\prime}(a_0) = \left[\frac{1-\frac{2a_0^2}{R^2}}{\sqrt{1-\frac{a^{2}_0}{R^{2}}}}
	-\frac{1-\frac{M}{a_0}}{\sqrt{1-\frac{2M}{a_0}+\frac{q^{2}}{a_0^{2}}}}
	\right].
	\end{equation}
	For a static shell, one can derive the inequality for $m_s^{\prime\prime}(a_0)$, which may be used for a stable configuration as the above expression containing two terms with opposite signs.
	
\begin{figure}[ht!]
		\includegraphics[width=0.45\textwidth]{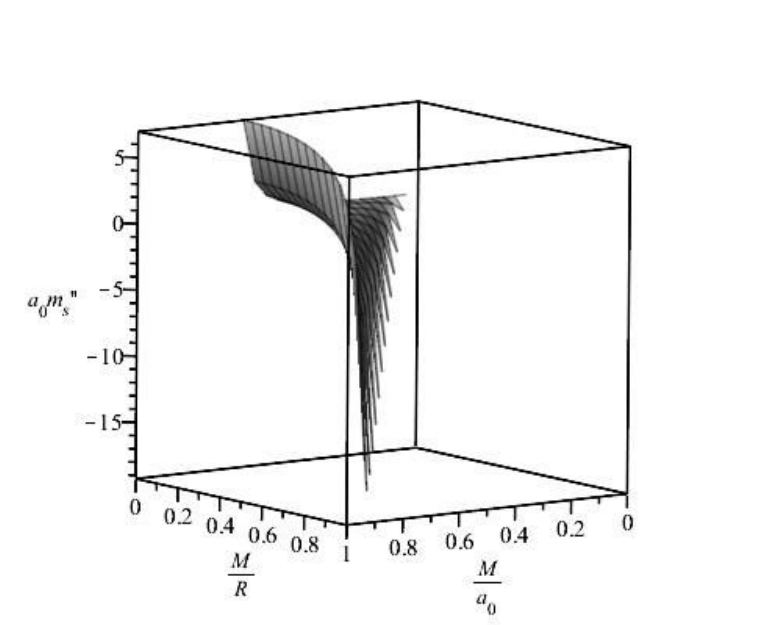} %
		\caption{\small \textit{Plot of the inequality (\ref{me}) for the function $a_0 m_s^{\prime\prime}$ in the case that $V = 0$. We define the graph for a positive surface energy density and for the values of $q/m < a_0$ and $a_0 < R$. The stability regions are given above the surface.} }\label{fig6}
	\end{figure}

	Generally, we require that the thin--shell matter satisfies the weak and dominant energy conditions. In this analysis, we shall adapt the cases of $\sigma > 0$, which correspond to positive surface energy densities (see the Appendix for more details). By considering a stable static solution at $a_0$, we must have:
	\begin{equation}\label{me}
	a_0 m_s^{\prime\prime}(a_0) \geq  \Bigg\{\frac{\left(\frac{a_0}{R}\right)^2\left[2\left(\frac{a_0}{R}\right)^2-3\right]}{\left(1-\frac{a^{2}_0}{R^{2}}\right)^{3/2}}
	-\frac{\left(\frac{M}{a_0}\right)^2 \Bigl[\left(\frac{q}{M}\right)^2 -1\Bigl]}{\left(1-\frac{2M}{a_0}+\frac{q^{2}}{a_0^{2}}\right)^{3/2}}
	\Bigg\}
	\end{equation}
	Now, from the master equation (\ref{me}), we mimic the stable equilibrium regions of the respective solutions.
	To determine the stability regions of this solution, we choose the parameters  such that the transition layer is located at some value between $|q| < a_0 < R$. Since the explicit form of the inequalities is extremely lengthy, we produce the graphical representation shown in Fig. \ref{fig6},	where the stability regions are represented above this surface.
	This is in good agreement with our previous results.

\section{Summary and Discussion}

	As an alternative to black holes, compact objects like gravastars have been proposed as a different final state of a gravitational collapse, though the evidence for the existence of black holes is well accepted astrophysically. In this paper, a spherically symmetric charged thin-shell gravastar has been investigated for a certain range of parameters where the metric potentials and electromagnetic fields are related in some particular relation called Guilfoyle's solutions.  We consider the gravastar composed of a de Sitter core, a thin shell, and an exterior Reissner-Nordstr\"om electrovacuum region. 
	
	The most relevant property is that within the $\delta$-shell models surface energy density and surface pressure are positive for a certain range of parameters. Therefore, the obtained solutions satisfy the NEC as illustrated in Fig. \ref{fig1}. For completeness, we have extended our analysis by exploring the linearized spherically symmetric radial perturbations about static equilibrium solutions. In this case, stability regions are given by the plot depicted in Figs.(\ref{fig3},\ref{fig4},\ref{fig5}). Furthermore, we have discussed the gravastar stability through the surface mass of the thin shell, and we have show that the obtained results are  in good agreement with our previous results. Therefore, we draw the conclusion that a variety of electrically charged gravastar solutions may be constructed resulting from the Guilfoyle exact solutions. 
    
Moreover, when considering charged stars it is useful to examine the mass-radius-charge bounds discussed so far in the literature. However, a separate study is needed for these quantities. Therefore, we plan in the near future to extend our work by considering a mass-radius-charge bounds discussed by Andreasson \cite{Andreasson:2008xw}, and Bohmer  \& Harko \cite{Boehmer:2007gq}.

	\subsection*{Acknowledgments}
	AB is thankful to the authority of	Inter-University Centre for Astronomy and Astrophysics, Pune, India for providing research facilities.
	
	\section*{Appendix: Static Gravastar}
	We derive the surface stress-energy tensor in terms of surface energy density $\sigma$ and surface pressure $\mathcal{P}$, around a stable solution situated at $a_0$. To see the qualitative behavior depicted in Fig. (\ref{fig1}-\ref{fig2}), let us start with the
	expression (59-60) and indroduce new varables $x = M/a_0$, and $y = M/R$, then the surface energy density and the surface pressure read, respectively
	
	\begin{eqnarray}
	\sigma (a_0) &=&  -\frac{1}{4\pi a_0}\left[\sqrt{1-\frac{2M}{a_0}+\frac{q^{2}}{a^{2}}}
	-\sqrt{1-\frac{a^{2}_0}{R^{2}}}\;\right]\\\nonumber
	&=& -\frac{1}{4\pi R}\frac{x}{y}\left[\sqrt{1-2x+\frac{q^{2}}{m^{2}}x^2}
	-\sqrt{1-\frac{y^2}{x^{2}}}\;\right]\,,
	\end{eqnarray}
	\begin{eqnarray}
	\mathcal{P}(a_0) &=&  \frac{1}{8\pi a_0}\left[\frac{1-\frac{M}{a_0}}{\sqrt{1-\frac{2M}{a_0}
			+\frac{q^{2}}{a^{2}_0}}}-\frac{1-\frac{2a^2_0}{R^{2}}}
	{\sqrt{1-\frac{a^{2}_0}{R^{2}}}}\right]  \\
	&=& \frac{1}{8\pi R}\frac{x}{y} \nonumber
	\left[\frac{1-x}{\sqrt{1-2x+\frac{q^{2}}{m^2}x^2}}-\frac{1-\frac{2y^2}{x^{2}}}
	{\sqrt{1-\frac{y^{2}}{x^{2}}}}\right].
	\end{eqnarray}
	
	Note that surface energy density and surface pressure are both positive in their respective range.

\end{document}